# An IoT-Based System: Big Urban Traffic Data Mining Through Airborne Pollutant Gases Analysis

Daniel. Firouzimagham, Mohammad. Sabouri, and Fatemeh. Adhami.

Applied Control & Robotics Research Laboratory, Shiraz University, Shiraz, Iran.
Daniel.firouzimagham@gmail.com, mohamadsabory@yahoo.com, Fatemeh.adhami@yahoo.com

*Abstract*— Nowadays, in developing countries including Iran, the number of vehicles is increasing due to growing population. This has recently led to waste time getting stuck in traffic, take more time for daily commute, and increase accidents. So it is necessary to control traffic congestion by traffic police officers, expand paths efficiently and choose the best way for decreasing the traffic by citizens. Therefore, it is important to have the knowledge of instant traffic in each lane. Todays, many traffic organization services such as traffic police officer and urban traffic control system use traffic cameras, inductive sensors, satellite images, radar sensors, ultrasonic technology and radio-frequency identification (RFID) for urban traffic diagnosis. But this method has some problems such as inefficiency in heavy traffic influenced by condition of the air and inability to detect parallel traffic. Our method suggested in this article detects traffic congestion based on IOT containing a smart system that gives us traffic congestion by calculating the air pollution amount in that area. According to conducted experiment, the results were satisfied.

## I. INTRODUCTION

According to 85% growth in population in the last two centuries, infrastructure developments are significant. Unfortunately, infrastructure developments of streets in some countries were not coordinated with the population growth rate; therefore, heavy traffic is observed in the level of urban pathways. This traffic congestion caused many problems such as city traffic. The most practical method in the city level to solve this problem is the use of a smart way for traffic measurement through optimizing traffic by recognizing the case traffic of each avenue. Available methods have some disadvantages that make them useless under some conditions. In this study, due to the inefficiency of previous methods measuring the traffic in some states, we are considering our new approach based on IoT. This system contains some sensors to measure air pollution, wireless data transfer system, and processor, which approximate traffic. In this research, we achieved significant accuracy using IoT and designing an array system and applying multi-point measurements through some feedback on the real amount of traffic.

### A. Urban transportation system introduction

In the process of measuring urban traffic, we use different methods in accordance with street capacity. The maximum traffic flow obtainable on a given roadway using all available lanes; usually expressed in vehicles per hour or vehicles per day is called street capacity.

Since the 1990s, specialists noticed that citizens occupy the streets' capacities quickly. Competition in producing comfortable and inexpensive car is decreasing road safety. On the other hand, technology development and internet expansion made an online connection between citizens and traffic management organizations easily; so developing a wise and objective management is available by the use of improving road network efficiency. Thus, intelligent transportation systems(ITS) were born in the 90s. ITS is a common term used for a set of communication technology, control, and processing data for transportation system. The primary function of ITS is to develop a routing choice level for each user. However, the other users lead to the overall traffic system improvements. ITS is a rule and order which improves safety and meets transportations needs all over the world.

### B. Transportation system management (TSM) is introduced for smart and efficient traffic controlling Traffic management with intelligent instruments in order to decrease traffic congestion in urban pathways is TSM.

Sensors and electronic, electromagnetic, photonic, and electro-mechanic systems are used in urban TSMs for a network of civic passages capacity measurement including:

- Electro mechanic sensors

This is a simple, familiar, and useful sensor in traffic detector systems. When a car crosses a speed bump (upgraded with electromechanical sensors), its mechanical key vibrates, and its processor detects the vehicle. By counting the number of cars in certain time, the traffic volume is given. Automatic depreciation in a long time and crossing at the same time problem are disadvantages of this method.

- Photonic sensors

In some systems, photonic sensors (lasers) are used to count cars and measure traffic volume. In this method, two sensors are fixed on both sides of street. A passing car tears the connection of these sensors, and the counter calculates the number of passing cars in a specific time. This system cannot count the parallel passing vehicles.

- Radar sensors

In this method, two sensors are fixed on both sides of street. A passing car tears the connection of these sensors, and the counter calculates the number of passing cars in a specific time. This system cannot count the parallel passing vehicles, low speed, or stopped vehicles in heavy traffics. This system works based on sending and achieving waves, so telecommunication devices affect the system.

- Technology-based on satellite information

Analyzing traffic congestion by satellite pictures is an expensive method; so it is used for long term traffic analysis, and it is useless for instant traffic analysis.

- RIFD technology

In this system, an RIFD sensor is fixed on each car and a certain place on the street, so the passing cars are detected. But we cannot count cars without the RIFD sensor in this method.

- Ultrasonic technology

The ultrasonic transceiver sends a wave. By analyzing the speed and sweep time of returned wave, speed, location, and a number of cars will be determined. The primary disability of this system is the low accuracy due to the different surfaces, density, and different materials because these will affect the returned wave quality.

- Traffic camera

This is the most common method in our country and all over the world. These cameras are fixed at particular angles on a street. It can calculate the traffic volume using the picture processing system, but in some certain viewing angle or specific lighting/atmospheric conditions can be useless.

A summary of previous method disabilities is given in the table below.

TABLE I
Comparison of Traffic Detection Systems

| number | name | explanation |
|---|---|---|
| 1 | Traffic camera | in some certain viewing angle or specific lighting/atmospheric conditions can be useless |
| 2 | Electro mechanic sensors | Mechanical depreciation in a long time and crossing at the same time problem |
| 3 | Technology-based on satellite information | Offline usage, failure in air pollution and night condition, expensive setup cost |
| 4 | Radar sensors | cannot count the parallel passing cars, low speed or stopped cars in heavy traffics |
| 5 | Ultrasonic technology | low accuracy due to the different surfaces, density, and different materials |
| 6 | RIFD technology | cannot count cars without RIFD sensor and parallel passing cars |
| 7 | Photonic sensors | Cannot count parallel passing cars and harmful air condition effect |

*C. Similar researches*

According to the importance of traffic and daily progress in making the cities intelligence based on IOT systems all over the world, extensive investigations were done on this issue.

In "a survey on urban traffic management system using wireless sensor network by Nellore, Lucian" [10] reference traffic measuring method using infrared sensors was described.

In some researches using the wave and echo for calculating the traffic volume method was reported. In this method, signal analysis is used.

In [12], photonic sensors were described, and in this research, we are using 3D transceivers for finding each car location.

In [6], we consider delays control that occurred at the over-saturated intersections. In this research, we are using CORSIM software for simulation. Researchers used a mathematic programming method to optimize timing planning at the intersections. We considered the controlling factors at the intersections, which is necessary for city traffic management, so city traffic researchers simonized dynamic network system traffic according to parking capacities, different car rout, and street types.

In Iran, a research was done using CORSIM software in Sanat square, Tehran.

## II. PROCEDURE DESCRIPTION

According to the importance of this issue, different methods were used such a Traffic camera
Electro mechanic sensors, Technology based on satellite information, Radar sensors, Ultrasonic technology, RIFD technology, Photonic sensors. However, we cannot use them under all conditions.

In this research, our purpose is to improve diagnosing congestion and volume of traffic methods and reduce above problems in the previous approaches by designing a system based on Internet of Things (IOT) and data mining using analyzing air pollution produced by vehicles. We created this system due to the effect of various conditions of temperature and humidity.

- Data Mining

The process of extracting the patterns from a large volume of raw data is called data mining. Data mining algorithms can extract information well if we pre-process on initial data as well as process on output data. The data mining process helps us to remove irrelevant and non-verbal data from our database and gives us useful information as well as accelerates the processes.[14-15]

First, we measured all the flue gasses. Then, we figured out some of the flue gasses have reduced dependency on flue gasses. Therefore, some experiment was done to detect the most relevant flue gasses that combustion take place in a car engine.

Measuring these kinds of pollution in environment test takes time; so we are unable to take sample of ambient air.

Therefore, in order to accelerate the measurements, we used electrochemical sensors. The output of these sensors gives us the concentration of air pollution in the testing area. The result of the experiment and observations shows us the level of air pollution in the testing area is relevant to the number of vehicles in the area and traffic congestion directly. This system works based on momentary measuring the concentration of air pollution caused by engine combustion in urban pathways.

Signals measured by sensors multiply with constants due to the pollution production rate. These numerical constants, calculated with experimental methods, show the relationship between each polluting gas and traffic congestion. So by processing these constants and sensors output, we can have a good approximation of traffic congestion. These constants are not errorless. The error occurs due to air condition, cultural, and social behaviors. To reduce the error effect, we used the given below methods.

- Real traffic congestion feedback

Our measurement accuracy is directly relevant to these constants, but they are time-variant constants; so it is vital to upgrade them according to the system's functional history.

- Array system installation

We installed a number of this system to minimum wind effect. Data was collected by wireless internet.

TABLE II
Device components

| Number | name | duty |
|---|---|---|
| 1 | processor | Processing each system's data |
| 2 | Power supply | Supplying each system's power |
| 3 | Wireless connection and network | Wireless data transfer |
| 4 | Unused hydrocarbon sensor | Detecting Unused hydrocarbon level |
| 5 | Smoke sensor | Detecting Smoke level |
| 6 | CO sensor | Detecting CO level |
| 7 | SO2 sensor | Detecting SO2 level |
| 8 | temperature and humidity sensor | Detecting temperature and humidity level |
| 9 | Wind speed sensor | Detecting Wind speed |

The reasons for choosing each polluting gas is given below:
- Sulfur dioxide ($SO_2$)

It is a polluting gas released from car exhaust not gas stoves, ovens and other vapor releasing sources and a large amount of exiting $SO_2$ in the area is caused by engine combustion of gasoline and petrol vehicles. Therefore, the amount of this polluting gas in the city plays a vital role in our new traffic congestion calculating system. In our researches, it was clear that we can diagnose traffic congestion by analyzing $SO_2$ concentration in the area. The level of $SO_2$ in the area can be related to traffic congestion with reasonable accuracy, so the more Sulfur dioxide concentration in the area, the more vehicles are detected. The amount of produced gas in each car varies due to the fuel quality and type of the vehicle, but this gas is detectable in all car exhausts, even in the lowest case. Thus, it is necessary to consider its concentration in the area.

- Unburned hydrocarbon ($H_xC_x$)

In the process of engine combustion of gasoline and petrol vehicles, a percentage of unburned fuel comes out of the exhaust. Dirty or low-quality car filters are determinant factors of this percentage. The amount of unburned hydrocarbons is related to the fuel quality, proper engine functioning, and environmental factors such as temperature and humidity, so the concentration of this gas in the testing area is determinative.

- Carbone monoxide (CO)

Carbone monoxide is another polluting gas caused by engine incomplete combustion and gas stoves, ovens, and other polluting gas releasing sources that are not relevant to traffic congestion; so in cold seasons, the appropriate constant to its amount decreases. This will reduce the effect of carbon monoxide on traffic congestions calculations.

- Soot

Due to the impurity of fuel, another chemical factor, and inefficient engine functions, carbon is produced. Therefore, by measuring its amount, the traffic congestion can be calculated by reasonable accuracy.

In this research, we did not pay attention to carbon dioxide's mount in the testing area because it is not relevant to the traffic congestion directly and might be a production of other polluting sources. If we entered this factor to our calculations, the output accuracy should have decreased.

To examine the system practically and test the accuracy, we chose Molasadra Street in Shiraz city. In a specific time, for a few days, the system's output was recorded. By array structure, the wind effect on gasses concentration was considered.

To increase the accuracy, devices were installed in 4 different parts of the street, and the mean of these four outputs was obtained. The system's feedback is possible enumeration in this case. This feedback is connected to the devices with IoT based connections.

The traffic detection algorithm was obtained from a large number of data set in 16 April to 3 September. In this calculation, the amount of pollutant gases caused by car fuel is measured each minute. Due to temperature and humidity, the amount of traffic is calculated. According to figures 1 to 3, the amount of traffic detected by the system is the lowest, and during the midday, the traffic of vehicles increases, so the number of pollutants grows more. Traffic is reduced between 12 and 16 hours, and the density of pollutants is reduced accordingly. Traffic rates start to rise since the early night. The most and the least polluting gas was Carbone monoxide and sulfur dioxide respectively.

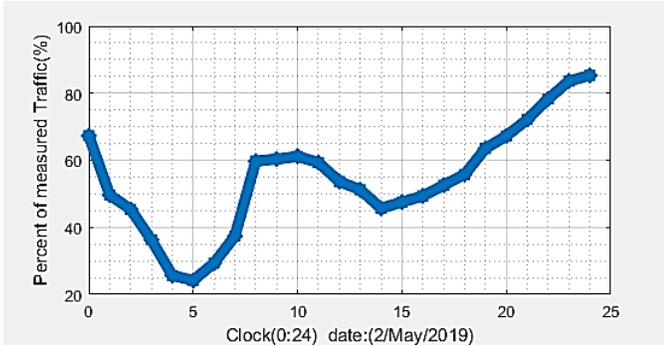
Fig. 1. The amount of traffic measured by the system in 2/May/2019

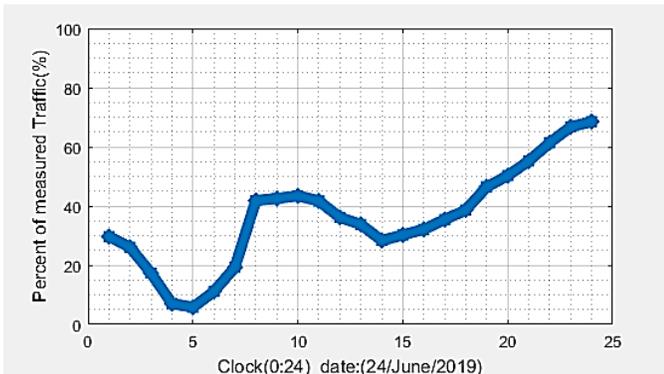
Fig. 2. The amount of traffic measured by the system in 24/June/2019

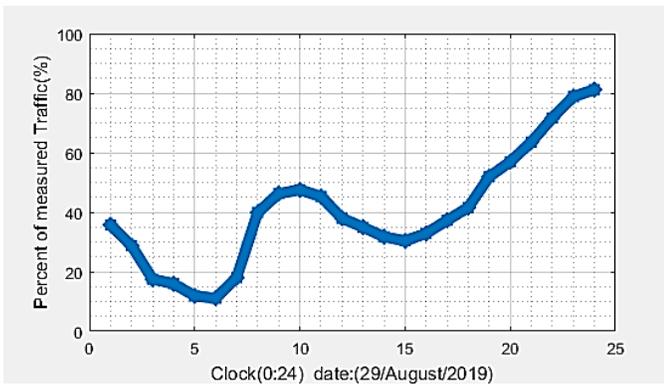
Fig. 3. The amount of traffic measured by the system in 29/August/2019

## III. Conclusion

According to the accomplished examinations and the results, we can state that a new method for diagnosing traffic congestion by analyzing air polluting gasses caused by vehicles is an almost efficient method which has solved the previous problems. This system can increase the output accuracy with IOT based on methods and present instant traffic congestion. While previous methods cannot give any information about traffic congestion and can only calculate traffic volume, this system can measure the traffic congestion in the street length using the information-sharing system based on IoT and using multi-devices to measure the traffic parameters. In this research, based on the practical examinations and comparing the system's output and results of enumerations, the presented method and this new system based on polluting gasses analysis are possible.